\documentclass{ws-procs975x65}

\begin{document}

\title{Relativistic Iron Lines in Galactic Black Holes: Recent Results
and Lines in the ASCA Archive}

\author{J. M. Miller}

\address{Harvard-Smithsonian Center for Astrophysics, 60 Garden Street (MS 67), Cambridge MA 02138, USA, jmmiller@cfa.harvard.edu~ (NSF Astronomy and Astrophysics Postdoctoral Fellow)}

\author{A. C. Fabian}

\address{Univ.\ of Cambridge Inst.\ of Astronomy, Madingley Road, Cambridge CB3 OHA, UK}

\author{M. A. Nowak,~~W. H. G. Lewin}

\address{MIT Center for Space Research, 70 Vassar Street, Cambridge MA
02139, USA}


\maketitle

\abstracts{Recent observations with {\it Chandra} and {\it
XMM-Newton}, aided by broad-band spectral coverage from {\it RXTE},
have revealed skewed relativistic iron emission lines in stellar-mass
Galactic black hole systems.  Such systems are excellent laboratories
for testing General relativity, and relativistic iron lines provide an
important tool for making such tests.  In this contribution to the
Proceedings of the 10th Annual Marcel Grossmann Meeting on General
Relativity, we briefly review recent developments and present initial
results from fits to archival {\it ASCA} observations of Galactic
black holes.  It stands to reason that relativistic effects, if real,
should be revealed in many systems (rather than just one or two); the
results of our archival work have borne-out this expectation.  The
{\it ASCA} spectra reveal skewed, relativistic lines in XTE
J1550$-$564, GRO~J1655$-$40, GRS~1915$+$105, and Cygnus X-1.}

\section{Background}

If the innermost regions of an accretion disk around a black hole are
irradiated by a source of hard X-rays, a characteristic ``disk
reflection'' spectrum is expected to result ([1], [2]).  These predictions have been borne-out observationally in
countless results.  The most prominent feature produced by reflection
is a fluorescent Fe~K$\alpha$ emission line.  Fe~K$\alpha$ emission
lines produced via reflection from the inner disk can serve as
extremely powerful probes of the environment closest to the black
hole, as the strong Doppler shifts and gravitational redshifts endemic
to the inner regions will be imprinted on the line profile ([3], [4]).
Relativistically-skewed lines and reflection spectra have been clearly
detected in a number of Seyfert-1 galaxies with {\it ASCA}, {\it
BeppoSAX}, and recently with {\it Chandra} and {\it XMM-Newton}.  The
much higher fluxes observed from stellar-mass Galactic black hole
systems prevented clear detection of such lines prior to {\it Chandra}
and {\it XMM-Newton}; new observations with these instruments have
revealed relativistic lines that are remarkably similar to those seen
in some AGN.  For reviews of these topics, the reader is referred to 
[5] and [6].

\section{Recent Developments}

The advanced spectrometers aboard {\it Chandra} and {\it XMM-Newton}
have revolutionized the study of relativistic Fe~K$\alpha$ emission
lines in stellar-mass Galactic black holes.  The line in Cygnus X-1
was revealed to consist of separate narrow and relativistic components
[7].  Most interesting of all is the potential for establishing black
hole spin using relativistic Fe~K$\alpha$ lines.  The innermost stable
circular orbit around a black hole with maximal spin is $r_{in} =
1.2~r_{g}$ (for $a=0.998$, where $a = cJ/GM^{2}$ and $r_{g} =
GM/c^{2}$), while that around a black hole with zero spin is $r_{in} =
6~r_{g}$.  Lines which are strongly skewed toward lower energies can
therefore indicate black hole spin.  The extremely skewed line profile
in the {\it XMM-Newton} spectrum of XTE~J1650$-$500 requires a black
hole with near maximal spin [8].  Observations of GX~339$-$4 with
{\it Chandra} [9] and {\it XMM-Newton} [10] have revealed
relativistic lines which strongly require a black hole with
near-maximal spin.  The {\it XMM-Newton} spectrum is particularly
exciting because the signal to noise achieved is comparable to the
best-defined relativistic lines seen in some Seyfert-1 galaxies.

The initial Cygnus X-1 result bolstered confidence among observers
that broad lines seen with low-resolution gas proportional counter
detectors, like those that flew aboard {\it BeppoSAX} and {\it ASCA},
and presently aboard {\it RXTE}, are robust features and can be
reliably fitted with advanced relativistic spectral models.  Analysis
of archival {\it BeppoSAX} spectra has revealed relativistic lines in
SAX~J1711.6$-$3808 [11], XTE~J1908$+$094 [12],
GRS~1915$+$105 [13], and most dramatically in XTE~J1650$-$500 [14].
Park et al.\ \cite{p04} fit relativistic line profiles to over 40 {\it
RXTE}/PCA observations of 4U~1543$-$475 during the bright phase of its
outburst, and found that skewed line profiles systematically provide
better fits than symmetric (Gaussian) line profiles during periods of
relatively strong hard X-ray emission.
 
\section{Analysis of Archival ASCA Observations}

We have undertaken a uniform analysis of archival {\it ASCA}/GIS
spectra of stellar-mass Galactic black holes in order to search for
relativistic Fe~K$\alpha$ emission lines, and herein we briefly
present the most exciting results from this effort.  While the {\it
ASCA}/SIS CCD spectrometer suffered from strong photon pile-up when
observing bright sources, the GIS gas spectrometer merely suffered
telemetry saturation without significant change to its spectral
response characteristics (T. Yaqoob, priv. comm.).  Even with
deteriorated detector response due to photon pile-up, hints of
extremely skewed emission line profiles can be seen in some SIS
spectra (see Figure 3 in [16], concerning SIS spectra of
GRO~J1655$-$40).

The {\it ASCA} spectra which appear in the HEASARC archive have been
filtered in a robust manner, making them robust for spectral analysis.
We have not attempted to extract background spectra for bright
Galactic sources; the source flux overwhelmingly dominates the
background, and the entire detector plane is illuminated by bright
sources.  For simplicity, herein we only report on fits to GIS-2
spectra; future work will include joint fits to GIS-2 and GIS-3
spectra.  The spectra were analyzed using XSPEC version 11.2.0.

In Table 1, we detail fits to four spectra from the following sources:
XTE J1550$-$564 (net exposure: 25.2~ksec, starting on 1998 Sept.\
23.9), GRO~J1655$-$40 (net exposure: 16.5~ksec, starting on 1994
Sept.\ 27.5) , GRS~1915$+$105 (net exposure: 65.4~ksec, starting on
2000 Apr.\ 21.5), and Cygnus X-1 (net exposure: 35.9~ksec, starting on
1996 May 30).  The ratio of the lines in these spectra to
phenomenological models is shown in Figure 1.

Skewed, relativistic lines are clearly detected in XTE J1550--564, GRO
J1655--40, and in Cygnus X-1.  The lines detected in XTE J1550$-$564
and GRO J1655$-$40 indicate that the black holes in these systems may
have a high degree of spin.  High frequency (few$\times 100$~Hz) QPOs
in these systems also suggest spin ([17], [18], [19]).  The evidence
for spin in Cygnus X-1 is less clear; future observations obtaining
better constraints will be essential to determine the nature of the
black hole in Cygnus X-1.  The best-fit physically-motivated models
for these spectra (see Table 1) are remarkably similar to those
applied to other Galactic black holes ([2], [3], [4]) and Seyfert-1
galaxies like MCG--6-30-15, Mrk~766, and NGC 4051.  This provides
evidence that the inner accretion flow geometry and emission
mechanisms may be very similar in these systems.  The line seen in
GRS~1915$+$105 is consistent with a radially recessed inner disk edge;
however, the constraints from this line are poor.  The fact that a
broadened line is detected in GRS~1915$+$105 is consistent with the
detection of a broad line in {\it BeppoSAX} spectra [13].

\section{Summary}

In the {\it Chandra} and {\it XMM-Newton} era, relativistic
Fe~K$\alpha$ emission lines have been revealed in stellar-mass
Galactic black holes.  They have proved to be robust diagnostics of
the innermost accretion flow region, which can also constrain the
nature of the central black hole.  Archival analyses of the kind
briefly detailed here show that relativistic lines are in fact common,
and not limited to a few special sources.  In the near future, we look
forward to the high effective area, spectral resolution, and broad
energy range of {\it ASTRO-E2}, to further refine studies of this
kind.  In the long run, {\it Constellation-X} will make it possible to
study changes in Fe~K$\alpha$ line profiles over short timescales;
such investigations will be especially powerful probes of the corona
--- disk interaction in Galactic black hole systems.

\begin{figure}[htbp]
\epsfxsize=8cm   %width of figure - will enlarge/reduce the figures
%\epsfbox{fig3.eps}
%\figurebox{2cm}{3cm}{} %to have a box alone 
\centerline{\epsfxsize=4.5in\epsfbox{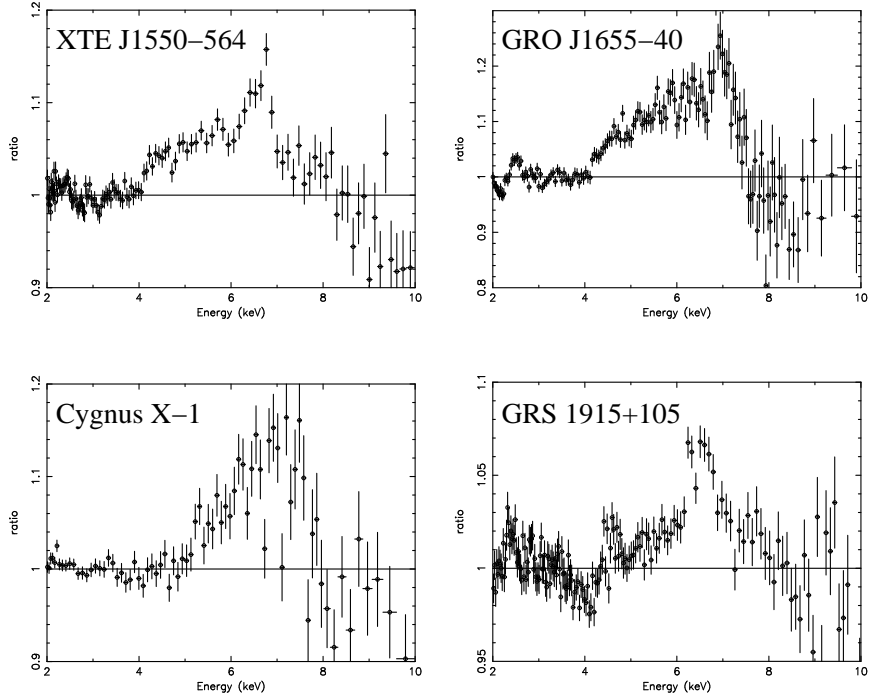}}   
\caption{Prominent relativistic line profiles in Galactic black holes
observed with {\it ASCA}.  The plots above show the ratio of GIS
spectra to a phenomenological continuum model consisting of multicolor
disk blackbody, power-law, and smeared edge
components.  A narrow Fe
XXV/XXVI absorption line and narrow Ca XX absorption line were fit to
GRO J1655$-$40 and GRS~1915$+$105.\label{asca_lines}}
\end{figure}
\eject

\begin{table}[htbp]
\tbl{Spectral fit parameters}
{\footnotesize
\begin{tabular}{@{}lllll@{}}
\hline
{} &{} &{} &{} &{}\\[-1.5ex]
{} & XTE~J1550$-$564 & GRO~J1655$-$40 & GRS~1915$+$105 & Cygnus X-1\\[1ex]
\hline
{} &{} &{phenomenological models} &{} &{}\\[1ex]
\hline
$N_{H}~(10^{21})$ & 0.7(1) & 1.3(1) & 4.0(1) & 5.3(3) \\[1ex]
$kT$~(keV) & $0.65_{-0.03}^{+0.7}$ & 1.26(1) & -- & 0.42(1) \\[1ex]
Norm. & $900^{+400}_{-200}$ & 150(10) & -- & 6100(500) \\[1ex]
$\Gamma$ & 2.2(2) & 5.2(3) & 2.24(8) & 2.6(1) \\[1ex]
Norm. & 2.3(4) & $5^{+8}_{-1}$ & $7.4^{+0.2}_{-0.6}$ & 1.1(1) \\[1ex]
$E_{Laor}$~(keV) & $6.4^{+0.6}$ & 6.8(2) & $6.40^{+0.6}$ & $6.8^{+0.2}_{-0.4}$  \\[1ex]
$q$ & $5^{+1}_{-2}$ & 5.5(5) & 5(1) & $4^{+2}_{-1}$ \\[1ex]
$r_{in}~(r_{g})$ & $2^{+2}_{-1}$ & $1.4^{+0.6}_{-0.2}$ & $1.8^{+1.5}_{-0.8}$ & $2.2\pm 1.0$ \\[1ex]
$i$~(deg) & $60^{+15}_{-30}$ & $45^{+15}_{-5}$ & $60^{+10}_{-20}$ & $45^{+15}_{-15}$ \\[1ex]
EW (eV) & 170(100) & 550(50) & $140^{+50}_{-60}$ & 400(80) \\[1ex]
Norm. ($10^{-3}$) & 17(10) & 57(4) & $18^{+6}_{-8}$  & $3.6\pm 0.8$ \\[1ex]
$E_{smedge}$ & 8.3(3) & 8.0 & 8.8 & $9.3_{-2.1}$ \\[1ex]
$\tau$ & $0.6^{+0.4}_{-0.2}$ & 1.0 & $0.3_{-0.2}^{+1.2}$ & $1.0_{-1.0}^{+0.5}$ \\[1ex]
$W_{smedge}$~(keV) & 7.0 & 7.0 & 7.0 & 7.0 \\[1ex]
$\chi^{2}/dof$ & 857.0/741 & 972.9/647 & 1113.8/751 & 780.8/654 \\[1ex]
\hline
{} &{} &{blurred reflection (pexriv) models} &{} &{}\\[1ex]
\hline
$N_{H}~(10^{21})$ & 0.58(2) & 1.3(2) & 4.1(!) & 4.8(2) \\[1ex]
$kT$~(keV) & 0.68(3) & 1.26(1) & -- & 0.45(1) \\[1ex]
Norm. & 1300(200) & 155(5) & -- & 5000(400) \\[1ex]
$\Gamma$ & $1.6_{-0.1}^{+0.3}$ & 6.0(1) & 2.30(5) & 2.4(3) \\[1ex]
Norm. & 0.5(1) & 10(1) & 7.8(6) & 0.6(4) \\[1ex]
$E_{Laor}$~(keV) & $6.5_{-0.1}^{+0.5}$ & $6.97_{-0.1}$ & $6.4^{+0.2}$ & 6.7(3) \\[1ex]
$q$ & 5(1) & 5.5(5) & $3^{+3}$ & $3.5_{-0.5}^{+1.0}$ \\[1ex]
$r_{in}~(r_{g})$ & $1.6_{-0.4}^{+4.0}$ & $1.8_{-0.6}^{+0.5}$ & $20_{-10}^{400}$ & $1.2^{+3.0}$ \\[1ex]
$i$~(deg) & 50(20) & 50(10) & 70(20) & 50(20) \\[1ex]
EW (eV) & 260(70) & 480(50) & 50 & 400(200) \\[1ex]
Norm. ($10^{-3}$) & 4.4(4) &{} & 4(3) & 2(1) \\[1ex]
$R$~($\Omega/2\pi$) & 1.0(3) & 1.0(2) & 0.3(2) & 1.0(3) \\[1ex]
$E_{fold}$~(keV) & 200 & 200 & 30 & 200 \\[1ex]
$\xi$ ($10^{4}$) & 1.0 & 1.0 & 0.1 & 1.0 \\[1ex]
$\chi^{2}/dof$ & 815.1/741 & 918.7/648 & 1041.6/752 & 849.5/655 \\[1ex]
\hline
\end{tabular}\label{table1} }
\begin{tabnote}
All errors are 90\% confidence errors.  Where errors are not given,
the parameter was fixed at the given value.  The Laor line energy
range was constrained to the 6.40--6.97~keV range, and the emissivity
index was constrained to vary in the 3--6 range.  Within pexriv, the
abundance flags were set to solar values, the inclination to the value
determined by the Laor line model, and the disk temperature to 1~keV.
The fit to the spectrum of GRO J1655$-$40 also included a Gaussian
absorption line (FWHM$=$0~keV, EW$=$25~eV) at 6.7~keV (Fe XXV).  The
spectrum of GRS~1915$+$105 included a Gaussian absorption line
(FWHM$=$0~keV, EW$=$8~eV) at 4.9~keV (Ca XX); the disk temperature in
pexriv was set to 0.1~keV for GRS~1915$+$105.\\ The extraordinary S/N
achieved with these observations makes residual narrow calibration
uncertainties important (particularly around 2 keV, and below), and
prevents formally acceptable fits in most cases.
\end{tabnote}
\vspace*{13pt}
\end{table}

\end{document}